\documentclass[useAMS,usenatbib]{mn2e}

\usepackage[T1]{fontenc}
\usepackage{aecompl}
\usepackage{graphicx}
\usepackage{gensymb}
\usepackage{epstopdf}
\usepackage{rotating}
\usepackage{lscape}
\usepackage{aas_macros}
\usepackage{hyperref}
\usepackage{xspace}
\usepackage{dsfont}
\usepackage{amsmath}
\usepackage{amsfonts}
\usepackage{amssymb}

\newcommand{\unit}[1]{\ensuremath{\, \mathrm{#1}}}

\newcommand\pccm{pc\,cm$^{-3}$\xspace}
\newcommand\halpha{H$\alpha$\xspace}
\newcommand\hbeta{H$\beta$\xspace}

\def\apjl{ApJ}                % Astrophysical Journal, Letters
\def\apjs{ApJS}               % Astrophysical Journal, Supplement
\def\aap{A\&A}                % Astronomy and Astrophysics
\begin{document}

\title[A Galactic Origin for FRB010621]{A Galactic Origin for the Fast Radio Burst FRB010621}
\author[K.W.\ Bannister et al.]{K. W. Bannister$^{1,2}$\thanks{E-mail:
keith.bannister@csiro.au}, G. J. Madsen$^{3,4,5}$\\
$^{1}$CSIRO Astronomy and Space Science, PO Box 76, Epping NSW 1710, Australia  \\
$^{2}$Bolton Fellow \\
$^{3}$Sydney Institute for Astronomy (SIfA), School of Physics, The University of Sydney, NSW 2006, Australia  \\
$^{4}$ARC Centre of Excellence for All-sky Astrophysics (CAASTRO) \\
$^{5}$Present address: Institute of Astronomy, University of Cambridge, Madingley Road, Cambridge CB3 0HA, UK \\
}

\date{Accepted 2014 January 29.  Received 2014 January 20; in original form 2013 October 10}
\pagerange{\pageref{firstpage}--\pageref{lastpage}} \pubyear{2013}

\maketitle

\label{firstpage}

\begin{abstract}
The recent detection of Fast Radio Bursts (FRBs) has generated strong interest in identifying the origin of these bright, non-repeating, highly dispersed pulses. The principal limitation in understanding the origin of these bursts is the lack of reliable distance estimates; their high dispersion measures imply that they may be at cosmological distances ($0.1 < z < 1.0$). Here we discuss new distance constraints to the FRB010621 (a.k.a J1852$-$08) first reported by Keane.
We use velocity resolved $H\alpha$ and $H\beta$ observations of diffuse ionised gas toward the burst to calculate an extinction-corrected emission measure along the line of sight. We combine this emission measure with models of Galactic rotation and of electron distribution to derive a 90\% probability of the pulse residing in the Galaxy.  However, we cannot differentiate between the two Galactic interpretations of Keane: a neutron star with unusual pulse amplitude distribution or Galactic black hole annihilation.
\end{abstract}


\section{Introduction}

In the past few years, there have been a growing number of  detections of unusual radio pulses from the Parkes radio telescope, known as Fast Radio Bursts (FRBs).
The six known FRBs \citep{lorimer2007bmr, Keane12,Thornton13}, are characterised by extremely high flux densities ($\sim$ Jy) over very short time scales (milli-seconds), high dispersion measures (DMs), and in one case, a scatter-broadened pulse profile. While there was for some time doubt as to whether these pulses are astronomical or terrestrial \citep{BurkeSpolaor11lg, Kocz12},  more recent discoveries support an astronomical interpretation \citep{Keane12, Thornton13}, in spite of the lack of multi wavelength counterpart.

It has not been definitively shown if the pulses are Galactic or extragalactic. Most of these discoveries favour an extragalactic origin ($0.1 \lesssim z \lesssim 1.0$), primarily because the DM substantially exceeds both the known dispersion measures from pulsars near to the line of sight, and / or a DM obtained from a commonly used Galactic model (e.g. NE2001 \citet{Cordes02}). If they are extragalactic, their extremely high luminosities can only be explained by exotic mechanisms such as annihilating black holes \citep{Rees77}, binary neutron star mergers \citep{Hansen01, Totani13} `combing' magnetic fields in expanding supernova shells \citep{Colgate71}, magnetar bursts \citep{Popov13}, collapse of supramassive neutron stars \citep{Falcke13} or gamma-ray bursts \citep{Zhang13, Bannister12}. There is considerable interest in identifying such classes of objects; they are also prime candidates as the electromagnetic counterparts to high frequency gravity wave sources that are expected to be discovered with upcoming facilities \citep{LIGO2012, Nissanke13}.

However, if fast radio bursts are Galactic, their characteristics can be attributed to somewhat less exotic mechanisms such as pulsar giant pulses with unusual pulse amplitude distributions \citep{Keane12}, or nearby flaring stars, where the large dispersion measure is explained by a blanket of ionised material around the star \citep{Loeb13}.

Because the progenitors of all of these highly-dispersed radio bursts are unknown, and because there are a small number of detections, it is important to examine each of the detections in detail. In this paper, we aim determine whether the burst named FRB010621 resides in the Galaxy or not.

FRB010621 (also known as J1852$-$08), was first reported by \citet{Keane11} and discussed in more detail by \citet{Keane12}. The burst position happens to lie toward an unusually low extinction region of the Galaxy just off the plane ($b=-4^\circ$).  Optical emission line spectroscopy of diffuse ionised gas has been studied extensively in this region \citep{Madsen05}. This fortunate coincidence allows us to use this spectroscopy to estimate the amount of Galactic ionised material in the direction of  FRB010621, and therefore determine the extent to which the DM of  FRB010621 (746 \pccm) can be attributed to ionised gas in our own Galaxy. The use of such spectroscopy has been used to successfully model distances to Galactic pulsars, albeit with less optimal dust corrections \citep{Schnitzeler12}. Our distance constraint will aid in the understanding the progenitor of  FRB010621 and in understanding statistical properties of other highly-dispersed radio bursts.

This paper is organised as follows: in \S\ref{sec:scutum} we describe existing observations of the low-extinction region of the ISM in which FRB010621 falls.  We discuss the emission measure toward the burst in \S\ref{sec:em} and outline a model that relates the EM to the DM in \S\ref{sec:dmcalc}.  In \S\ref{sec:distance} we describe our constraints on the distance to the burst. We discuss the implications of our results on the origin of FRB010621 in \S\ref{sec:discussion} and summarise our results in \S\ref{sec:conclusion}.

\section{Scutum star cloud: a low-extinction window toward the inner Galaxy}
\label{sec:scutum}

\subsection{Background}

The Scutum star `cloud' is an optically bright, $\sim 5^\circ$ diameter region of unusually high apparent stellar density centred on $(l, b) = (27^\circ, -3^\circ)$.  The brightness is attributed to its proximity to the Galactic plane and the relatively low extinction along the line of sight \citep[][and references therein]{Madsen05}.  Optical emission line spectroscopy of diffuse ionised gas toward this cloud shows emission out to large kinematic distances. Two spectra taken near the location of the radio burst are shown in Figure \ref{fig: spectra}. In these directions, diffuse interstellar gas that follows Galactic rotation exhibits a maximum radial velocity (in the local standard of rest reference frame) equal to the tangent point velocity $v_{LSR} = +113 \unit{km s^{-1}}$; this corresponds to a kinematic distance of $D_{\odot} \approx$ 8 kpc
for a flat rotation curve assuming the current IAU recommended values for the rotational velocity $\Omega$ = 220 km s$^{-1}$ and distance to the Galactic centre $R_\odot$ = 8.5 kpc.

\subsection{\halpha\ and \hbeta\ spectra and kinematics}
The \halpha\ and \hbeta\ spectra shown in Figure \ref{fig: spectra} were obtained with the Wisconsin H-Alpha Mapper \citep[WHAM;][]{Haffner03}.   WHAM measures the average surface brightness of emission over a 1$^\circ$ circular diameter beam with a spectral resolution of $\approx$  12 km~s$^{-1}$.
Under typical interstellar conditions, photoionised gas will produce \halpha and \hbeta photons with a ratio of \halpha/\hbeta = 3.9 \citep{Osterbrock06}. However, interstellar dust absorbs and scatters the bluer \hbeta\ photons more than \halpha\ photons.  \citet{Madsen05} measured the departure of \halpha/\hbeta\ from the expected value to quantify the amount of interstellar extinction as a function of kinematic distance.  They found that interstellar dust produced an optical depth at \halpha\ of $\tau_{H\alpha} \approx 2$ ($A_V \approx$ 3 mag) out to the tangent point distance of the Scutum cloud (see bottom row Figure \ref{fig: spectra}).

The spectra show three strong peaks of emission near $v_{LSR} = 0 \unit{km~s^{-1}}$, +50~$\unit{km ~s^{-1}}$ and +100~$\unit{km~s^{-1}}$; these velocities correspond to distances to emission from the solar neighbourhood, the Sagittarius spiral arm, and the Scutum spiral arm, respectively.  The weak emission seen beyond the tangent point velocity could be due to broadening of the line centred near the velocity of the Scutum arm. The broadening may arise from a combination of thermal broadening of the lines ($\approx 10\unit{km~s^{-1}}$ at $T_e ~ 10^4$~K) and non-thermal broadening such as turbulence, bulk motion, and/or non-circular motion in the inner Galaxy \citep{Madsen05}
In addition, some of the highest velocity emission could be associated with gas near the tangent point but with a velocity in excess of that associated with circular rotation (e.g., \citet{Lockman02}).

\subsection{The position of FRB010621}
The radio burst FRB010621 was detected in one 14\arcmin\ diameter beam of the Parkes telescope, centred on
$(l, b) = (25.4342^\circ, -4.0042^\circ)$ \citep{Keane12}.
The two WHAM observations of the Scutum cloud that are closest to the burst are centred on $(l, b) = (25.08^\circ, -4.25^\circ)$ and $(25.52^\circ, -3.40^\circ)$.

The overlap of the WHAM and Parkes beams implies that the burst could be in one (or neither) of the WHAM beams. If the position of the burst is equally likely anywhere within the Parkes beam, then there is a 84\% chance that the burst is in the ``main'' WHAM beam toward $(l, b) = (25.08^\circ, -4.25^\circ)$, a 0.7\% chance of lying in the alternate WHAM beam, and a 15\% chance of not lying in either WHAM beam.  As there is such a small chance of the burst lying in the alternate WHAM beam, we will ignore it for the remainder of this paper. We also note that the variation in H$\alpha$ surface brightness from diffuse ionised gas does not vary substantially on 1$^\circ$ angular scales \citep{Haffner03}.

\section{Emission measure toward FRB010621}
\label{sec:em}

\begin{figure}
\includegraphics[scale=0.35]{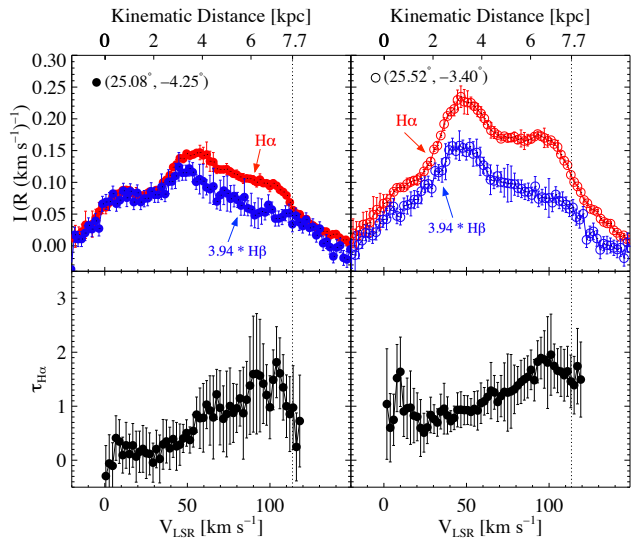}
\caption{\halpha\ (red) and \hbeta\ (blue) WHAM spectra of diffuse ionised gas toward the radio burst (top row) in two \protect $1^\circ$ diameter beams nearest  FRB010621. The \hbeta\ surface brightness is multiplied by 3.94, the expected ratio of \halpha/\hbeta in the absence of extinction. The bottom row shows the estimated optical depth at \halpha. The left and right columns show data for the two WHAM lines of sight closest to the location of the burst.  The kinematic distance, based on a flat rotation curve, is shown along the top. The tangent point distance, marked with a vertical dotted line, is about 7.7~kpc in this direction. The burst was detected at $l=25^\circ.43$, $b=-4^\circ$. (Figure is adapted from \citet{Madsen05}).}\label{fig: spectra}
\end{figure}

In the absence of extinction, a column of recombining electrons with emission measure EM will generate \halpha\ emission with a surface brightness $I_{H\alpha}$ given by

\begin{equation}
{\rm{EM}} \equiv \int_{0}^{\infty} n_e(l)^2 dl = 2.75 T_4^{0.9} I_{H\alpha}
\end{equation}
where $n_e(l)$ is the electron density in units of cm$^{-3}$, $l$ is the distance along the line of sight in units of pc,
$T_4$ is the electron temperature of the gas in units of $10^4$ K, and $I_{H\alpha}$ is measured in rayleighs\footnotemark\footnotetext{1 R $= 10^6/4\pi$ photons~cm$^{-2}$s$^{-1}$sr$^{-1} = 2.4 \times 10^{-7}$ ergs~cm$^{-2}$s$^{-1}$sr$^{-1}$ at \halpha.}\citep{Haffner03}.
The observed values of $I_{H\alpha}$ toward the two WHAM beams shown in Figure \ref{fig: spectra} on the left and right panels are 13.6 $\pm$ 1.2 R and 20.8 $\pm$ 1.5 R, respectively.  These values are calculated by integrating the spectra over the full velocity range and taking into account the uncertainties in the continuum level.

In order to convert the observed values of $I_{H\alpha}$ to the emission measure (EM), we need to apply a dust correction.  However, this correction (derived from \halpha/\hbeta) suffers from a number of uncertainties (see \S4 of \citet{Madsen05} for a complete discussion), principally due to the unknown distribution of the dust and gas.  For the purposes of illustration, we consider two models for the dust distribution, representing the two extremes of how the dust and ionised gas are mixed: a foreground screen model and model where the gas and dust are well-mixed. In both cases, we assume the optical depth of the dust at $H\alpha$ is $\tau = 2$.

For the foreground screen model, the observed $I_{H\alpha}$ should be increased by a factor of $e^{\tau} = 7.4$, while for the well-mixed model the correction factor is only $\tau / (1 - e^{-\tau}) = 2.3$.   Assuming a temperature of  $T_4 = 1.0$ \citep{Madsen05} the extinction corrected emission measure toward the ``main'' WHAM beam is $276\pm24$ cm$^{-6}$~pc assuming the foreground screen model; it is $86\pm7$ cm$^{-6}$~pc assuming the well-mixed model.

Figure \ref{fig: spectra} shows that the values of $\tau_{H\alpha}$ increase sharply at a kinematic distance of $\approx$ 3-4 kpc, suggesting there is a screen of dust at that distance, which strongly favours a foreground screen model. Due to the uncertainty in the dust geometry and the uncertainty in the location of the burst in the WHAM beams, we adopt a value for the extinction corrected EM toward FRB010621 of 276 cm$^{-6}$~pc.

\section{Simple model for emission and dispersion measure}
\label{sec:dmcalc}

In order to use the EM and DM to constrain the distance to the radio burst, we need to know how the free electrons are distributed along the line of sight.  We adopt a simplified model in which the electrons reside in clouds of mean density $n_0$; the fraction of the line of sight occupied by the clouds is defined as the filling factor ($f$). We define the heliocentric distance to the radio pulse to be $D_p$.
We define the heliocentric distance over which Galactic electrons are contributing significantly to the EM to be $D_{H\alpha}$.
In principle, the emission measure derived from $I_{H\alpha}$ is sensitive to all recombining electrons along the line of sight. However, the dominant contribution to the EM toward FRB010621 is from electrons in the Galaxy; we neglect any contribution from the intergalactic medium where $n_e$ is much less than the average Galactic mid-plane density of $\approx$ 0.1 cm$^{-3}$ \citep{Inoue04}.   It then follows that DM = $n_0 f D_p$ and EM = $n_0^2 f D_{H\alpha}$. With this parameterisation, the distance to the pulse is
\begin{equation}
D_p = \frac{\rm{DM}}{\sqrt{{\rm{EM}}}} \frac{\sqrt{D_{H\alpha}}}{\sqrt{f}}. \label{eq:dp}
\end{equation}

Using DM = 746 pc cm$^{-3}$, EM = 276 cm$^{-6}$pc, $D^{10}_{H\alpha} = D_{H\alpha}/10$ kpc, and $f_{0.1} = f/0.1$, this can be expressed as
\begin{equation}
D_p = 13.6 \frac{\sqrt{D^{10}_{H\alpha}}}{\sqrt{f_{0.1}}} \rm{kpc}.
\end{equation}

The assumptions implicit in this model are that the filing factor ($f$) and the mean electron density $n_0$ have the same values for the distances $D_p$ and $D_{H\alpha}$ and that the clouds occupying the line of sight have uniform density $n_0$. Observations of pulsar scintillation have shown that the medium is not uniformly dense. Consequently the filling factor in Eqn.~\ref{eq:dp} could be larger by some factor proportional to the magnitude of the density fluctuations. A larger filling factor implies a smaller pulse distance. By adopting a uniform density assumption, the distance $D_p$ in Eqn.~\ref{eq:dp} should be considered an upper limit.

This simple model has been successfully combined with observations of pulsars to infer structural parameters such as the scale height of diffuse ionised gas in the Galaxy \citep{Reynolds91, Berkhuijsen06, Gaensler08, Schnitzeler12}.  Note that those analyses are largely based on nearby pulsars with $D_p \ll D_{H\alpha}$; those observations of EM therefore require some correction to account for the different path lengths. That problem is mitigated by only considering high latitude pulsars (where $D_p \approx D_{H\alpha}$) or by adopting another model for how EM varies in the Galaxy. Such modelling is outside the scope of this paper.

An HII region along the line of sight toward FRB010621 could be responsible for large density fluctuations that would invalidate our model. However, a large angular size, nearby HII region would appear in our data as an \halpha\ enhancement at a particular velocity, and we detect none here (Fig.~\ref{fig: spectra}). Furthermore, there is no catalogued HII regions \citep{Sharpless59, Blitz82, Fich84} that cross this line of sight, and radio recombination line searches \citep[e.g.][]{Caswell87} have not surveyed this line of sight.

The existence of HII regions along the line of sight does not conclusively show that the pulse is Galactic, as we have no way of knowing whether the pulse was behind a given HII region or not. On the other hand, if we can show that there is sufficient \emph{diffuse} ionised gas along the line of sight, the presence of a foreground HII region would only serve to reduce the pulse distance.

The simple model for the EM and DM described above has some shortcomings that make it difficult to estimate a precise distance to the FRB.  However, this model is sufficient to infer whether the  FRB is Galactic or extragalactic.

\section{Is FRB010621 Galactic or Extragalactic?}
\label{sec:distance}

According to our simple model, the distance to the pulse depends on the filling factor $f$, the path length of the \halpha\ emitting gas $D_{H\alpha}$ and the dust distribution.  We use a Bayesian approach to quantify the distance to the pulse and the uncertainty.
In order to determine a posterior distribution on the probability to the distance to the pulse, we have performed a simple Monte-Carlo simulation, assuming prior distributions for the parameters $f$, $D_{H\alpha}$, and dust model. In the following sections we discuss the priors on each parameter and the results of the simulations. Note that with a latitude of $-4.0^\circ$, the radio burst is 0.6~kpc and 1.6~kpc below the Galactic plane for values of $D_p$ of 8~kpc and 24~kpc, respectively.

\subsection{Prior on $H\alpha$ distance ($D_{H\alpha}$)}
Figure \ref{fig: spectra} shows that we detect \halpha\ emission out to the tangent point velocity. This places a lower limit on $D_{H\alpha}$ of $\approx$ 8 kpc.  If there is \halpha\ emitting gas beyond this distance, it will appear at radial velocities lower than $v_{LSR} \approx +100$ km/s.

We place an upper limit on $D_{H\alpha}$ by considering the radial size of the Galaxy.   \citet{Cordes02} used dispersion measures toward more than 1,000 pulsars to construct a popular three dimensional model for the distribution of free electrons.  The model contains a thick disk component that is truncated at a Galactocentric distance of 17 kpc. This thick disk component is the dominant contributor to the total electron column density in the outer parts of the Galaxy. At a Galactic longitude of 25$^\circ$ and assuming a cylindrical geometry with $R_{\odot} = 8.5$ kpc, their model implies that $D_{H\alpha} < 24$ kpc.  The actual value is likely to be less than 24 kpc because this does not account for the fact that the radio burst is not in the mid plane ($b = -4^\circ$) and the scale height of $n_e$ is only 1-2 kpc \citep{Gaensler08, Savage09}.

In the intergalactic medium (IGM) beyond the thick disk, $n_e \simeq 2.1 \times 10^{-7} \unit{cm^{-3}}$ \citep{Inoue04}, which is a factor of $\sim 10^{-6}$ lower than the average Galactic mid-plane density. Because $EM\propto n_e^2$, the contribution of IGM to the measured EM can be safely ignored, and the upper limit to $D_{H\alpha}$ is robustly constrained by the size of the Galaxy.

Given these two limits, we assume a uniformly distributed prior on $D_{H\alpha}$ between 8 and 24~kpc.

\subsection{Prior on filling factor ($f$)}
The observational limits on values for $f$ are less well constrained.
As shown in equation \ref{eq:dp}, the filling factor can be measured directly using dispersion measures of pulsars which have reliable distances (e.g., from parallax).
According to the latest version of the ATNF Pulsar Catalogue\footnote{\url{http://www.atnf.csiro.au/people/pulsar/psrcat/}} \citep[version 1.47;][]{Manchester05}, there are 88 pulsars within 5$^\circ$ of FRB010621. However, none of them have a reliable, model independent distance.

\citet{Reynolds91} was the first to unambiguously measure $f$ by comparing the dispersion and emission measures toward pulsars in high latitude globular clusters that are well off the Galactic plane. They found that $f \gtrsim 0.2$ with marginal evidence that $f$ increased with vertical height $|z|$.
\citet{Gaensler08} analysed the distribution of DM and EM  for 51 pulsars with known, reliable distances over a range of values for $|z|$.  They fit their data to a model with a filling factor that increases exponentially from $f = 0.04\pm$0.01 in the mid-plane up to $f = 0.3$ at a vertical height of $|z| \approx$ 1--1.5 kpc.
\citet{Berkhuijsen08} conducted an independent analysis of similar data for 34 pulsars with reliable distance measurements. They found that $f = 0.08\pm$0.02 at the mid-plane, but did not determine robustly how, or if, $f$ changes with $|z|$.

These results are consistent with the traditional picture of the ISM. In the mid-plane, the ISM is largely comprised of cold neutral and molecular gas \citep{Kulkarni87}.  At higher altitudes, warm diffuse ionised and neutral gas are the principal phases.  Note that with a latitude of -4.0$^\circ$, the radio burst is 0.6 kpc and 1.6 kpc off the Galactic plane for values of $D_p$ of 8 kpc and 24 kpc, respectively.

The measured values of $f$ throughout the literature are generally bounded by $0.03 < f < 0.3$. We therefore assume a uniformly distributed prior on $f$ between 0.03 and 0.3.

\subsection{Prior on dust mixing model}

As discussed in \S\ref{sec:em}, the magnitude of the extinction correction depends on the (unknown) distribution of interstellar dust along the line of sight.  We considered two models: a foreground screen model and a model where the gas and dust are well mixed.  Our adopted value of the extinction corrected EM toward FRB010621 (276 cm$^{-6}$pc) corresponds to a dust extinction correction factor that is more consistent with a foreground screen model. However, in order to quantify this uncertainty, we consider three scenarios in our Monte-Carlo simulations.  In one scenario we assume the foreground screen model, and in another scenario we assume a well-mixed model. In the third scenario, called the `alternating' scenario, half of our realisations use a foreground screen model, and the other half use a well-mixed model.

\subsection{The pulse distance}

\begin{figure}
\centering
\hspace{-2cm}
\includegraphics[width=1.2\linewidth]{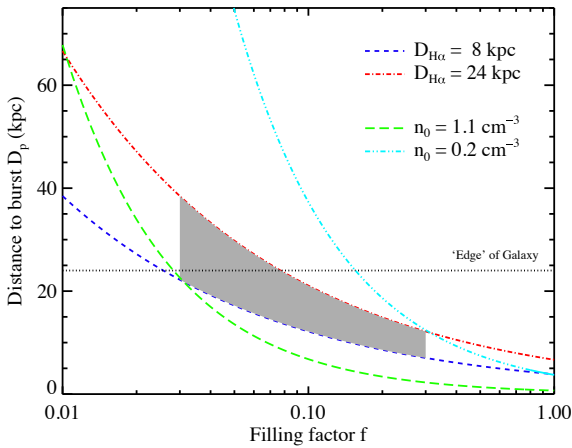}
\caption{\label{fig: dist} Distance to the burst as a function of filling factor of the ionised gas using an extinction corrected EM of 276 cm$^{-6}$pc. The grey shaded area shows the range of possible values of $D_p$ bounded by 8 kpc $< D_{H\alpha} <$ 24 kpc and $0.03 < f < 0.30$. The black horizontal dotted line is placed at the canonical `edge' of the Galaxy along the line of sight to the burst (24 kpc). The lower and upper limits on the average electron density over the corresponding range of values of $D_p$ are $0.2 < n < 1.1$; lines with these values of constant density are shown in aqua and green, respectively.}
\end{figure}

The full range of plausible values for the distance to the radio burst (using Equation \ref{eq:dp} and assuming an extinction corrected EM value of 276 cm$^{-6}$ pc) are shown in Figure \ref{fig: dist}.  The Figure shows that the distance $D_p$ is bounded by values between 7~kpc and 40~kpc. The implied average electron density over this distance range is $0.2~\unit{cm^{-3}} < n_0 < 1.1~\unit{cm^{-3}}$. We note, however, that for values of $D_{H\alpha}$ greater than 24~kpc, our model comprising a single value for $n_0$ and $f$ is no longer valid.

\begin{figure}
\centering
\includegraphics[width=\linewidth]{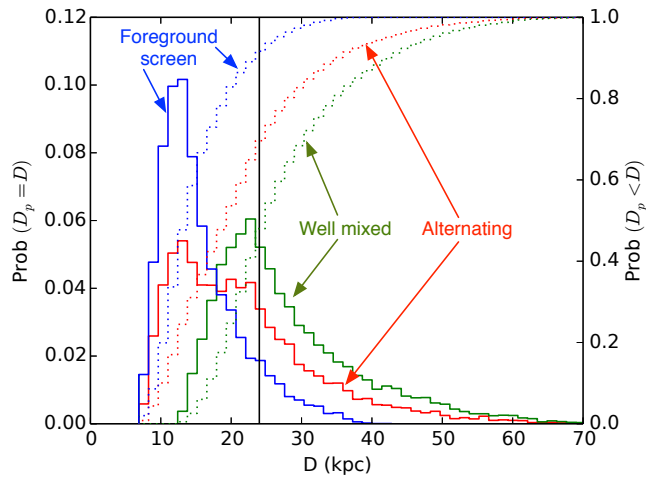}
\caption{\label{fig:dhist} Histograms (solid lines) and cumulative histograms (dotted lines) of pulse distances from $10^4$ realisations of the prior distributions described in Section \ref{sec:distance} for the line of sight to FRB010621. The x-axis represents the range of distances ($D$) considered by the models. The y-axis on the left hand side of the panel shows the posterior probability for $D_p = D$; the results are shown as solid coloured lines. The y-axis on the right represents the cumulative probability for $D_p < D$; the results are shown as dotted coloured lines. The solid black vertical dotted line is placed at the canonical `edge' of the Galaxy along the line of sight to the burst (24 kpc).}.
\end{figure}

The results of our Monte Carlo simulations are shown in Figure~\ref{fig:dhist}.
We created $10^4$ realisations from the prior distribution on $f$ and $D_{H\alpha}$, assuming Gaussian distributed EMs derived in section \ref{sec:em} for the two standard dust models, and the `alternating' model. The posterior probability distributions on the pulse distance are shown in Fig.~\ref{fig:dhist}. The median pulse distance (and 1$\sigma$ width) is $14\pm6$ kpc, $25 \pm 10$~kpc, and $19 \pm 11$~kpc for the foreground screen (blue), well mixed (green) and alternating (red) models, respectively. The pulse distance is less than the distance to the `edge' of the Galaxy ($D_{H\alpha}< 24$ kpc) in 90\%, 68\% and 45\% of realisations for the models, respectively.

\section{Discussion}
\label{sec:discussion}

\subsection{Implications of a Galactic origin}

Assuming the foreground screen dust model, we conclude that FRB010621 is Galactic (i.e. $D_p < 24\unit{kpc}$) with 90\% confidence. \citet{Keane12} outline two possible explanations for a Galactic origin: pulsed, sporadic emission from a neutron star (e.g. a rotating radio transient \citep{Mclaughlin09}) or an annihilating Galactic black hole. While our analysis cannot conclusively differentiate between either scenario, we encourage the community to consider these Galactic options more seriously. In particular, it may be that FRB010621 is a neutron star whose pulsed emission has only been observed in this one burst, in spite of 15.5~hr of additional follow-up \citep{Keane12}.

We note that our EM measurement is consistent with the measured pulse broadening due to scattering. \citet{Keane12} puts an upper limit on the scatter broadening as $< 3$~ms. Assuming the relations for diffracting scattering from \citet{Cordes98}, and a upper limit distance of 22~kpc, we obtain a pulse broadening time of $\simeq 2.4$~ms, which is consistent with the measured limit.

Our results show there is a 10\% probability that FRB010621 is extragalactic. We therefore cannot provide new constraints on an extragalactic distance, beyond those in the original claims of \citet{Keane12}. We note that our conclusions have no bearing on the distances of other bursts, in particular those at higher Galactic latitudes \citep{Thornton13, BurkeSpolaor13}.

\subsection{The absence of high-DM pulsars along the line of sight}

The ATNF pulsar catalogue lists 8 pulsars within a $2^{o}$ radius of FRB010621, with the largest DM being 388~\pccm \citep[J1846$-$0749]{Lorimer06}. The absence of pulsars along this line of sight with DMs similar to the DM of FRB010621 could be seen as evidence for a low Milky Way DM along this line of sight, and therefore an extragalactic origin of FRB010621.

However we note that this absence of pulsars is more likely due to the fact that most pulsars are  found in the mid-plane of the Galaxy. The known pulsar population has been shown to be exponentially distributed with height from the mid-plane, with a scale-height of 300~pc \citep{Faucher-Giguere06}. With this scale height, at a latitude of $-4^\circ$, the density of pulsars drops to 10\% of the mid-plane density at a distance of 10~kpc. It is not surprising therefore, that we find no pulsars with very high DMs at this latitude, as there are simply very few pulsars to be found so far from the mid-plane. We note that this argument applies only to the distribution of catalogued pulsars associated with the Milky Way. The sorts of objects that may be the source of FRB010621 may be distributed very differently.

\subsection{Limitations of Galactic electron models}

The combination of \halpha\ and \hbeta\ observations and a low extinction region of the Galaxy where the rotation velocity is directly related to the distance provides a uniquely favourable measurement of the pulse distance at a relatively low latitude. Nonetheless, the uncertainty on our estimate of the pulse distance is still large. As \citep[Section 4.2]{Cordes02} themselves point out, there are substantial errors in the NE2001 model, and workers in the FRB field should be aware of them.

\section{Conclusions}
\label{sec:conclusion}

We have taken advantage of the fortuitous location of the fast radio burst FRB010621 toward a low-extinction region of the Galaxy to estimate the extinction corrected emission measure toward the burst. We have combined this with the observed DM to the pulse and evaluated the pulse distance and DM to the edge of the Galaxy, and their uncertainties, with a Monte Carlo simulation. Under the assumption that the interstellar dust toward the pulse acts as a foreground screen, we find  that FRB010621 is located within the Galaxy with 90\% confidence and the distance is $14\pm6$~kpc. While our results strongly favour a Galactic interpretation, we cannot differentiate between the two Galactic scenarios proposed by Keane: namely a pulsar with unusual amplitude distribution, or annihilating Galactic black hole.  Our results also show the limitations of using the Galactic electron models, such as NE2001 to estimate distances to dispersed pulses, especially near the Galactic plane.

We note that while FRB010621 is likely to reside in the Galaxy, other FRBs may well be extragalactic. The other FRBs are at much higher Galactic latitude where the electron column density is not large enough to account for the observed DMs by much larger factors.  It therefore could be that FRB010621 belongs to a different class of FRBs. There are a number of archival and new observational surveys that are searching for FRBs \citep{Macquart10, Hassall13, Trott13}.  We anticipate that the discovery of more of these kinds of bursts will significantly increase our limited understanding of this strange phenomenon.

\section{Acknowledgements}
The authors would like to thank Simon Johnston, Matthew Bailes and Evan Keane, and the anonymous referee for their valuable comments and suggestions on this manuscript. Parts of this research were supported by the Australian Research Council Centre of Excellence for All-sky Astrophysics (CAASTRO), through project number CE110001020.

\bibliographystyle{mn2e}
\bibliography{Master.bib}

\label{lastpage}
\end{document}